# A preliminary study of asymmetric vocal fold vibrations: modeling and *"in-vitro"* validation.

**Nicolas Ruty[1], Annemie Van Hirtum[1], Xavier Pelorson[1], Avraham Hirschberg[2], Ines Lopez[3]**

[1]Institut de la Communication Parlée – UMR5009 – CNRS/INPG/Université Stendhal - Avenue Félix Viallet, 38031 Grenoble Cedex 01 - France

[2]Eindhoven University of Technology, Applied Physics, Gas Dynamics and Aero-acoustics 5600 MB Eindhoven, The Netherlands

[3]Eindhoven University of Technology Mechanical Engineering, Dynamic and Control, PO Box 513, WH 0.134, 5600 MB Eindhoven, The Netherlands

nicolas.ruty@icp.inpg.fr, xavier.pelorson@icp.inpg.fr, a.hirschberg@tue.nl, i.lopez@tue.nl

***Abstract.*** *This paper deals with some of aspects of the influence of asymmetry on vocal folds vibrations. A theoretical model of vocal fold asymmetry is presented. The influence of asymmetry is quantitatively examined in terms of oscillation frequency and pressure threshold. The theoretical model is compared to "in-vitro" experiment on a deformable replica of vocal folds. It is found that asymmetry strongly influences the oscillation subglottal pressure threshold. Moreover, the vocal fold with the highest mechanical resonance frequency imposes the oscillation fundamental frequency. The influence of geometrical asymmetry instead of purely mechanical asymmetry is shown.*

## 1. Introduction

Voiced sound production can be disturbed in case of vocal fold pathology. A consequence of pathologies is commonly irregular or asymmetric vibrations of the vocal folds. Voice quality is altered and voiced sounds are difficult to produce. In term of voice quality, additional in-vivo evidence is presented by Mergell et al. (2000), using high-speed imaging of the vocal fold during oscillations. The use of an asymmetry factor to quantify the influence of asymmetry between the two vocal fold has been proposed by Steinecke and Herzel (1995). A more global approach has been proposed by Schwarz et al. (2004), taking into account not only the asymmetry factor Q but also other parameters like the subglottal pressure Ps. Then using an inversion procedure, it was proposed to quantify all the parameters of a modified two mass model. In this paper, we propose an asymmetric model of the vocal folds, based on a modification of the two mass model described by Lous et al. (1998). To test its validity and to quantify the effects of asymmetry, an experimental set-up based on an asymmetric deformable replica of the vocal folds is used.

## 2. Theoretical modeling

The model used is derived from the symmetrical two-mass vocal fold model introduced by Lous et al. (1998). Two symmetrical coupled oscillators represent each vocal fold. Each oscillator is defined by the following parameters: m/2 is the mass (m is the mass of a vocal fold), k is the spring stiffness, $k_c$ is the coupling spring stiffness, and r is the damping. In each vocal fold, the set of parameters can be different. Consequently, the model of vocal fold can be asymmetric (e.g.: figure 1).

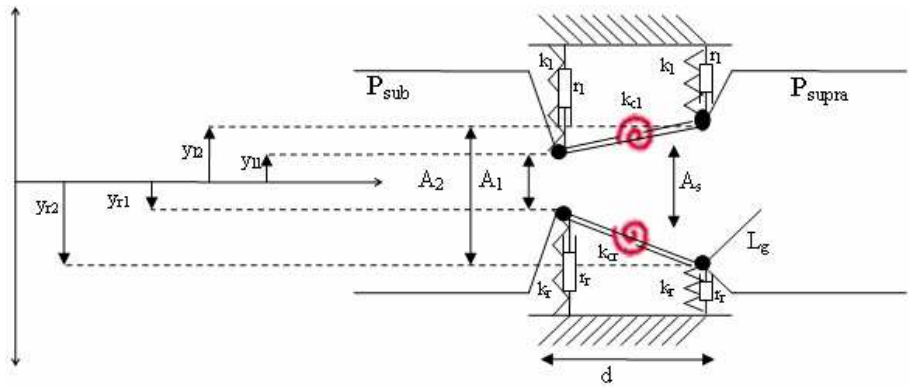

**Figure 1.** An asymmetric two-mass model of vocal folds. $P_{sub}$ is the subglottal pressure. $P_{supra}$ is the supraglottal pressure. $L_g$ is the glottal width. d is the length of glottis. $A_1$, $A_2$, $A_s$ are the cross-sectional areas (masses, airflow separation point). $y_{l1}$, $y_{l2}$, $y_{r1}$, $y_{r2}$ are the vertical positions of the masses. $k_l$ and $k_r$ are the spring stiffnesses. $k_{cl}$ and $k_{cr}$ are the coupling spring stiffnesses. $r_l$ and $r_r$ are the damping constants.

As a result, there are two mechanical equations for each vocal fold:

$$\begin{cases} \dfrac{m_i}{2} \dfrac{\partial^2 y_{i1}(t)}{\partial t^2} = -k_i(y_{i1}(t) - y_{i10}) - k_{ci}(y_{i1}(t) - y_{i10} - y_{i2}(t) + y_{i20}) - r_i \dfrac{\partial y_{i1}(t)}{\partial t} + F_{i1} \\ \dfrac{m_i}{2} \dfrac{\partial^2 y_{i2}(t)}{\partial t} = -k_i(y_{i2}(t) - y_{i20}) - k_{ci}(y_{i2}(t) - y_{i20} - y_{i1}(t) + y_{i10}) - r_i \dfrac{\partial y_{i2}(t)}{\partial t} + F_{i2} \end{cases} \quad (1)$$

where $m_i$ is the vibrating mass of a vocal fold (i={l,r}), $y_{i1}$ and $y_{i2}$ are the displacement of each mass compared to the axis of symmetry of the glottis, $y_{i10}$ and $y_{i20}$ are the rest position of each mass, $F_{i1}$ and $F_{i2}$ are the external pressure.

The asymmetry of this mechanical model is commonly expressed by the parameter Q, used for example by Steinecke and Herzel (1993) or Mergell et al. (2000):

$$\begin{aligned} k_r &= Qk_l \\ m_r &= m_l/Q \end{aligned} \quad (2)$$

One can also take into account the asymmetry of coupling stiffness with a parameter $Q_c=k_r/k_l$ (Schwarz et al., 2004), and the geometrical asymmetry of rest position vocal fold.

This mechanical model is coupled to an airflow description. Through the glottis geometry (figure 1), the airflow is modeled by a Poiseuille flow and a moving separation point is assumed (ad-hoc criterion). Thus, one can write the pressure distribution through the glottis (3) and the volume flow velocity (4):

$$P_{sub} - P(x,t) = \frac{1}{2}\rho U_g^2 \left(\frac{1}{A^2(x,t)}\right) + 12\mu L_g^2 U_g \int_{x_0}^{x} \frac{dx}{A^3(x,t)}, \text{ if } x < x_s \qquad (3)$$

$$P(x,t) = P_{supra}, \text{ if } x > x_s,$$

$$U_g = \frac{-12\mu L_g^2 \int_{x_0}^{x_s} \frac{dx}{A^3(x,t)} + \sqrt{\left(12\mu L_g^2 \int_{x_0}^{x_s} \frac{dx}{A^3(x,t)}\right)^2 + 2(P_{sub} - P_{supra})\rho\left(1/A_s^2\right)}}{\rho\left(1/A_s^2\right)} \qquad (4)$$

where $P_{sub}$ and $P_{supra}$ are the sub and supra glottal pressures, $P(x,t)$ is the pressure distribution, $\rho$ is the air density, $U_g$ is the volume flow velocity, $\mu$ is the viscosity coefficient, $L_g$ is the glottal width, $A(x,t) = L_g(y_l(x,t) + y_r(x,t))$ indicates the glottal cross-sectional area, $A_s = 1.2\min(A(x,t))$ is the cross-sectional area at the separation point $x_s$.

Due to the flow description, $F_{i1}$ and $F_{i2}$ depend on a lot of parameters ($F_{i,j\in\{1,2\}} = F_{i,j\in\{1,2\}}(y_{l1}, y_{l2}, y_{r1}, y_{r2}, P_{sub}, P_{supra})$). This system of four equations is mechanical asymmetric if the two vocal folds have different set of control parameters (spring stiffnesses, damping constant), and can be geometrical asymmetric (rest positions of each masses).

This aeromechanical model is coupled with the vocal tract, which is assimilated to a single degree of freedom system. It is described by the following equation:

$$\frac{\partial^2 \psi(t)}{\partial t^2} + \frac{\omega_A}{Q_A}\frac{\partial \psi(t)}{\partial t} + \omega_A^2 \psi(t) = \frac{Z_A \omega_A}{Q_A} u \qquad (5)$$

where $\frac{\partial \psi(t)}{\partial t} = p$, with p the acoustic pressure at the entrance of the vocal tract, $\omega_A$ is a resonance pulsation, $Q_A$ is the quality factor of this resonance, $Z_A$ is the peak value of impedance at resonance $\omega_A$, u is the acoustic airflow velocity.

In this paper, the theoretical model is analyzed thanks to two relevant quantities in voiced-sound production: the phonation on-set pressure threshold (subglottal pressure to sustain oscillations) and the fundamental frequency of oscillations.

## 3. Experimental set-up

The aim of this study is to test the validity of the theoretical model. For reasons of control and reproducibility, a mechanical replica of the vocal folds is used. It consists in two metal pieces covered with latex (figure 2, [e]). These two pieces are encapsulated into a rigid structure (figure 2, [f]). Each vocal fold replica is filled with water, thanks to a water reservoir (figure 2, [g] and [h]).

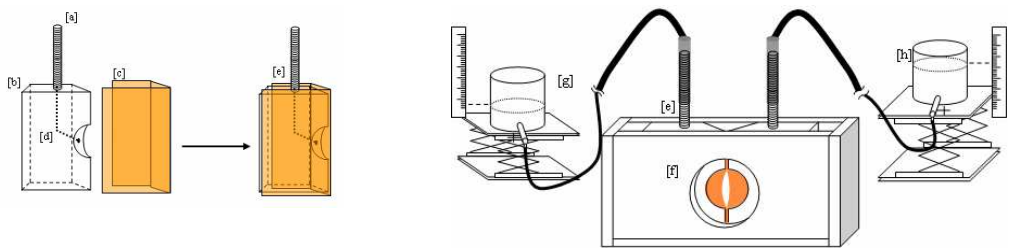

**Figure 2.** Deformable replica of the vocal folds made of metal, covered with latex, and filled with water. [a] screw (hollowed out) for water supply. [b] rigid part of a vocal fold replica. [c] deformable part of a replica, made of latex. [d] channel for water supply. [e] a vocal fold replica (rigid part + deformable part). [f] rigid structure maintaining the two vocal fold replicas. [g], [h] reservoir for the control of water pressure in each vocal fold replica.

The pressure (internal pressure Pc of a vocal fold replica) of water influences the mechanical characteristics of the replica. For example, when the pressure is increased, by increasing the height of the water column, the replica is tensed. Moreover, the internal pressure can be imposed different for each vocal fold replica. Hence the replica is asymmetric. The replica is coupled to a pressure reservoir and to a downstream resonator of 17 cm length (figure 3) as described by Ruty et al (2007). The reservoir is fed by a compressor. The upstream pressure (up to 4000 Pa) is controlled by a Norgren regulator type 11-818-987. Thus, airflow can be forced through the vocal folds replica. Under certain conditions, self-sustained oscillations appear. The characteristics of the oscillations depend on the upstream pressure, on the internal pressure Pc of the two "vocal fold", and consequently on the asymmetry between the two "vocal folds".

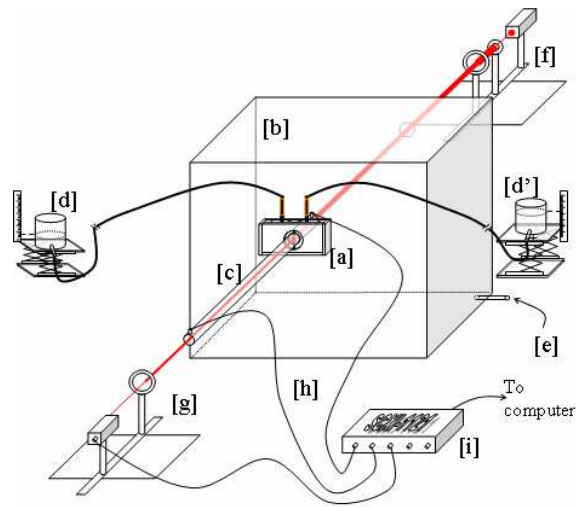

**Figure 3.** Experimental set-up producing flow induced vibration of the vocal fold replica. [a] vocal fold replica. [b] pressure reservoir. [c] downstream resonator of 17 cm length. [d], [d'] water reservoirs controlling internal pressure of each vocal fold replica. [e] air supply. [f] laser diode, optical lenses to increases the laser beam width. [g] photodiode, optical lens to focalize the beam on the sensor. [h] pressure sensors. [i] National Instruments amplifier SCXI-1121

Several physical quantities can be measured on this experimental set-up. The movements of the replica are measured by way of an optical set-up. It is composed of a laser diode, three convergent lenses (f=100 mm and f=50 mm), and of a photodiode BPW34. Pressure measurements are performed by way of two pressure sensors (Kulite XCS-0.93-0.35-BAR-G). The first one, placed just before the replica, allows to measure the upstream pressure. The second is placed at the end of the downstream resonator. A measurement microphone Bruël and Kjaer is used to measure acoustic pressure at the end of the resonator.

Electrical signals from the sensors are filtered and amplified with a National Instruments SCXI 1121 board. This board is connected to the computer by way of a National Instruments BNC 2110 input/output card and a National Instruments PCI-6225 acquisition card.

For calibration of the sensors, one proceeds as following. The pressure sensors are calibrated using a manometer. The photodiode is calibrated using rectangular apertures with known dimensions of 0.01 mm precision. The measurement microphone is calibrated using a Bruël and Kjaer Sound Level Calibrator Type 4230, which produce a pure sinusoidal audio signal of 1000Hz frequency and 94dB amplitude.

## 4. Results and discussion

In this section, we examine the influence of vocal fold asymmetry on two relevant quantities for voiced sound production: the oscillation pressure threshold and the fundamental frequency of oscillations. A stability analysis is performed on the theoretical model. Then, mechanical parameters (masses, spring stiffness, and damping) are extracted from the frequency response of the deformable replica. Finally, experimental data and theoretical prediction are compared and discussed.

### 4.1. Stability Analysis

The theoretical model is examined around equilibrium. Each variable is considered as the sum of an equilibrium value and a fluctuation. For example $y_{1r} = \bar{y}_{1r} + y'_{r1}$, where $\bar{y}_{1r}$ is the equilibrium value and $y'_{r1}$ is the fluctuation. Thus, equations (1) and (5) are written as following with (i={l,r}):

$$\begin{cases} \dfrac{m_l}{2}\dfrac{\partial^2 y'_{l1}}{\partial t^2} = -k_l y'_{l1} - k_{cl}(y'_{l1} - y'_{l2}) - r_l \dfrac{\partial y'_{l1}}{\partial t} + \dfrac{\partial F_{l1}}{\partial y_{l1}} y'_{l1} + \dfrac{\partial F_{l1}}{\partial y_{l2}} y'_{l2} + \dfrac{\partial F_{l1}}{\partial y_{r1}} y'_{r1} + \dfrac{\partial F_{l1}}{\partial y_{r2}} y'_{r2} + \dfrac{\partial F_{l1}}{\partial P_{\sup ra}} p \\ \dfrac{m_l}{2}\dfrac{\partial^2 y'_{l2}}{\partial t} = -k_l y'_{l2} - k_{cl}(y'_{l2} - y'_{l1}) - r_l \dfrac{\partial y'_{l2}}{\partial t} + \dfrac{\partial F_{l2}}{\partial y_{l1}} y'_{l1} + \dfrac{\partial F_{l2}}{\partial y_{l2}} y'_{l2} + \dfrac{\partial F_{l2}}{\partial y_{r1}} y'_{r1} + \dfrac{\partial F_{l2}}{\partial y_{r2}} y'_{r2} + \dfrac{\partial F_{l2}}{\partial P_{\sup ra}} p \end{cases} \quad (6)$$

$$\dfrac{\partial^2 \psi(t)}{\partial t^2} + \dfrac{\omega_A}{Q_A}\dfrac{\partial \psi(t)}{\partial t} + \omega_A^2 \psi(t) = \dfrac{Z_A \omega_A}{Q_A}\left(\dfrac{\partial U_g}{\partial y_{l1}} y'_{l1} + \dfrac{\partial U_g}{\partial y_{l2}} y'_{l2} + \dfrac{\partial U_g}{\partial y_{r1}} y'_{r1} + \dfrac{\partial U_g}{\partial y_{r2}} y'_{r2} + \dfrac{\partial U_g}{\partial P_{\sup ra}} p\right) \quad (7)$$

This system can be written using a state-space representation $\dot{x} = Mx$ where the state-space vector is

$$x = \left[ y_{l1} \quad y_{l2} \quad y_{r1} \quad y_{r2} \quad \psi \quad \dfrac{\partial y_{l1}}{\partial t} \quad \dfrac{\partial y_{l2}}{\partial t} \quad \dfrac{\partial y_{r1}}{\partial t} \quad \dfrac{\partial y_{r2}}{\partial t} \quad \dfrac{\partial \psi}{\partial t} \right] \quad (8)$$

Then, for a given initial geometry and a set of mechanical and acoustical parameter, the subglottal pressure $P_{sub}$ is varied from 0 to 5000 Pa. For each subglottal pressure, the eigenvalues of the state-space matrix are calculated. The sign of the real part of the eigenvalues is analyzed. If the real part is positive, that involves the presence of oscillation. The fundamental frequency of the oscillation is determined by $f0=Im(\lambda)/2\pi$, where $\lambda$ is an eigenvalue ($Re(\lambda)>0$). In some case, two or three eigenvalues have a positive real part. That involves the presence of more than one oscillation regime.

### 4.2. Extraction of the model parameters

As described in Ruty et al. (2007), the link between empirical parameters and the model parameters has to be known. For some parameters, the relationship is evident (e.g. geometrical parameters $L_g$, d, subglottal pressure $P_{sub}$). Other parameters (spring stiffness, damping constant, and masses) need to be estimated.

The estimation is made using the experimental set-up described in section 3. For a given internal pressure, the frequency response is measured as described by Ruty et al. (2005). The parameters k and r are obtained using the following relationship:

$$\omega_0 = \sqrt{\frac{2k}{m}}, \quad Q_0 = \frac{m\omega_0}{2r} \tag{9}$$

where $\omega_0$ is a resonance pulsation and $Q_0$ is the quality factor of the resonance.

The mass is estimated with the relationship:

$$m = \frac{L_g d Q_0}{\omega_0^2 Z_0} \tag{10}$$

where $Z_0$ is the amplitude of the measured resonance peak.

The coupling spring stiffness is obtained using another resonance of the frequency response ($\omega_1 > \omega_0$):

$$k_c = \alpha k, \text{with } \alpha = \frac{1}{2}\left[\left(\frac{\omega_1}{\omega_0}\right)^2 - 1\right] \tag{11}$$

### 4.3. Comparison between experimental data and theoretical prediction

Using the experimental set-up described in section 3, the influence of asymmetry is analyzed. An internal pressure of 4000 Pa in the left vocal fold replica is imposed. Internal pressure of the right vocal fold replica is varied from 1500 Pa to 3000 Pa by step of 500 Pa, and from 3000 Pa to 7500 Pa by step of 250 Pa. For each internal pressure in the right vocal replica, the same protocol is followed. The frequency response of the replica is measured and the corresponding mechanical parameters are obtained. Then, the upstream pressure is increased up to 5000 Pa. Different oscillation regimes appear as the upstream pressure increases. The measured on-set pressure thresholds and the fundamental frequencies of these oscillation regimes are recapitulated and compared with the theoretical predictions in figure 4.

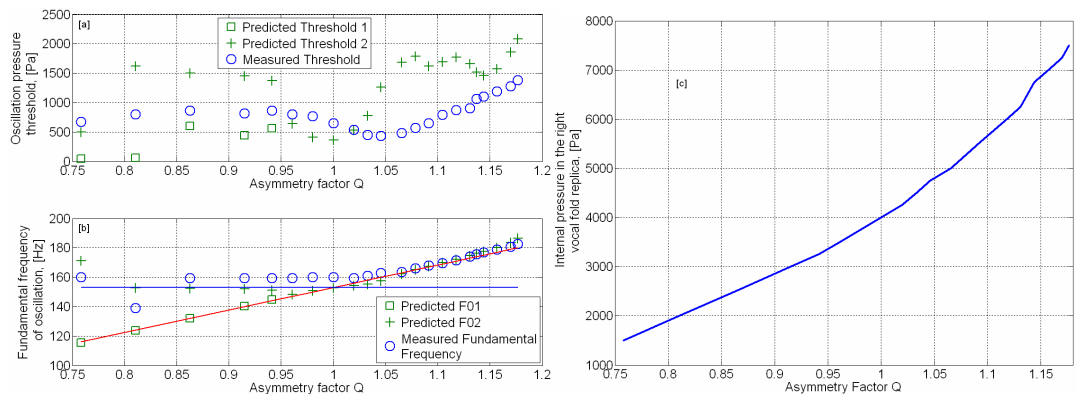

**Figure 4.** Comparison between experimental data and theoretical predictions. [a] oscillation pressure thresholds are plotted as a function of the asymmetry factor $Q=f_{right}/f_{left}$, where $f_{right}$ and $f_{left}$ are the measured resonance frequencies of each vocal fold replica. [b] experimental and predicted frequency of oscillation as a function of the asymmetry factor Q. Full lines in blue and red indicate the mechanical resonance frequencies of each vocal fold replica (blue=left, red=right). [c] Internal pressure in the left vocal fold replica as a function of the asymmetry factor.

### 4.4. Discussion

Concerning the pressure threshold and oscillation fundamental frequency, one can note that the influence of asymmetry is strong. Hence, the experimental pressure thresholds are increased when Q is not close to 1. This increase is different according to whether Q is higher or lower than 1. When Q>1, the pressure increase is clear and seems to be the consequence of the higher internal pressure in the right vocal fold. When Q<1, oscillations seem to be 'controlled' by the left vocal fold replica, of which internal pressure is higher. Theoretical predictions are qualitatively in agreement with the experimental data. When Q>1, the predicted thresholds increase, but quantitatively more than in experimental data. When Q<1, two oscillation regimes are predicted by the theoretical model. The frequency $F_0$ of each regime corresponds to the mechanical resonance frequency of a vocal fold replica (left or right), as see figure 4b. When Q>1, only one regime remains. Its frequency is close to the mechanical resonance of the right vocal fold. Experimentally, the same phenomenon is observed, the frequency of oscillations is close to highest mechanical resonance between the two replicas.

Finally, one can note that the influence of asymmetry is different depending on the factor Q, more precisely if Q<1 or Q>1. Indeed, we have deliberately chosen to define $Q=f_{right}/f_{left}$ while Steinecke and Herzel (1995) have chosen to define Q as the ratio of the lower and higher frequency ($Q=f_{low}/f_{high}$). The reason of our choice is that in our case, the values of Q correspond to an internal pressure of the vocal fold replica. The variation of internal pressure influences not only the mechanical resonance frequency, but also the initial aperture. As shown by Van Hirtum et al. (2006) initial condition has an influence on oscillations. That's why one needs to take into account not only mechanical asymmetry but also geometrical asymmetry, or more precisely the asymmetry of the rest position of each vocal fold.

# 5. Conclusion

An asymmetric model of the vocal folds has been proposed. Asymmetry is taken into account using the asymmetry factor Q, but in addition using all the mechanical and geometrical parameters. The theoretical model is compared to experimental data obtained with an asymmetric vocal fold replica. Comparison between theoretical prediction and experimental data leads to the following conclusions:

- Asymmetry increases the oscillation pressure threshold. The increase depends on the factor Q, but also on the initial geometry.

- Experimentally, fundamental frequency of oscillations is close to the highest mechanical resonance frequency between the vocal fold replicas. Theoretically, this result is well predicted when Q>1, whereas for Q<1 two oscillation regimes are predicted corresponding to the mechanical resonance of each vocal fold.

- The single asymmetry factor Q cannot explain the complexity of asymmetry. One needs to take into account all the mechanical and geometrical parameters.